\documentclass[10pt,conference]{IEEEtran}
\usepackage{color}
\usepackage{graphicx}
\usepackage{amsmath}
\usepackage{amsthm}
\usepackage{mathtools}
\usepackage{makecell}
\usepackage{bm}
\usepackage{hyperref}
\usepackage{multirow}
\usepackage{amsmath}
\usepackage{amssymb}
\usepackage{amsthm}
\usepackage{graphicx}
\usepackage{float}
\usepackage{listings}
\usepackage{xcolor}
\usepackage{subcaption}
\usepackage{tikz}
\usepackage[ruled, linesnumbered]{algorithm2e}
\usepackage{algpseudocode}
\usepackage{algcompatible}

\newcommand{\cG}{{\mathcal G}}

\usepackage{tikz} 
\usetikzlibrary{matrix,positioning}
\usepackage{amsfonts}
\usepackage{amssymb}
\hypersetup{
   colorlinks = true,
    urlcolor=blue
}
\begin{document}

\markboth{Mostafa Atallah}
{A mixed-integer program for circuit execution time minimization with precedence constraints}

\title{A mixed-integer program for circuit execution time minimization with precedence constraints}

 \author{\IEEEauthorblockN{Mostafa Atallah\IEEEauthorrefmark{1}\IEEEauthorrefmark{2}, James Ostrowski\IEEEauthorrefmark{1}, Rebekah Herrman\IEEEauthorrefmark{1}\IEEEauthorrefmark{3}
 }
 \IEEEauthorblockA{\IEEEauthorrefmark{1}Industrial and Systems Engineering, University of Tennessee at Knoxville, Knoxville, TN, USA \\
 \IEEEauthorrefmark{2}Department of Physics, Faculty of Science, Cairo University, Giza 12613, Egypt}
 
 Email: \IEEEauthorrefmark{3}rherrma2@utk.edu}

\maketitle
\begin{abstract}
    We present a mixed-integer programming (MIP) model for scheduling quantum circuits to minimize execution time. Our approach maximizes parallelism by allowing non-overlapping gates (those acting on distinct qubits) to execute simultaneously. Our methods apply to general circuits with precedence constraints. First, we derive closed-formulas for the execution time of circuits generated by ma-QAOA on star graphs for a layered, greedy, and MIP schedules. We then compare the MIP schedule against layered and greedy scheduling approaches on the circuits generated by ma-QAOA for solving the MaxCut problem on all non-isomorphic connected graphs with 3-7 vertices. These experiments demonstrate that the MIP scheduler consistently results in shorter circuit execution times than greedy and layered approaches, with up to 24\% savings.
\end{abstract}

\section{Introduction}
\label{sec:intro}

Quantum computing has emerged as a promising paradigm for solving complex optimization problems \cite{shaydulin2024evidence, montanaro2024quantum, srinivasan2018efficient}, but realizing its potential requires efficient circuit scheduling. A fundamental challenge in quantum computing is the mapping of logical quantum circuits onto physical hardware while respecting architectural constraints and optimizing for performance metrics such as circuit depth and gate count. Integer programming techniques have proven particularly effective in addressing these challenges.

Recent work has demonstrated the power of mixed-integer programming (MIP) for quantum circuit scheduling and optimization. Nannicini et al. \cite{nannicini2022optimal} developed a MIP model for optimally mapping logical quantum circuits onto hardware with limited connectivity, while Bhoumik et al. \cite{Bhoumik2024ResourceawareSO} extended this approach to schedule multiple quantum circuits simultaneously while maintaining nearest-neighbor qubit layout and minimizing qubit wastage, which refers to reducing the number of unused qubits on the hardware device. The challenge of satisfying nearest-neighbor constraints, which require interacting qubits to be physically adjacent, has been specifically addressed through MIP formulations that minimize the number of required SWAP gates \cite{boccia2019swap, bringman2024mathematical}, with related work by de Souza et al. \cite{de2019finding} focusing on optimizing CNOT gate counts for IBM's quantum architectures. 

The complexity of quantum circuit scheduling has led to diverse mathematical programming approaches. Orenstein et al. \cite{orenstein2024quantum} proposed using binary integer nonlinear programming for resource allocation and qubit mapping, while Itoko et al. \cite{itoko2020scheduling} demonstrated the effectiveness of constraint programming for improving circuit execution time. Bhattacharjee et al. \cite{Bhattacharjee2017DepthOptimalQC} introduced a MIP-based approach that optimizes quantum circuit depth while handling arbitrary qubit topologies, moving beyond the limitations of linear nearest-neighbor (LNN) architectures. For quantum annealing applications, Bernal et al. \cite{bernal2020integer} developed integer programming solutions for the minor-embedding problem. There exist more specialized MIP models for gate scheduling for circuits that have precedence constraints between gates, where certain quantum gates must be completed before others can begin \cite{Alam2020CircuitCM, alam2020efficient, tan2020optimal, guerreschi2018two, Lechner2018QuantumAO}.

In this work, we develop a new MIP model for scheduling quantum circuits where gates may have different execution times and must respect precedence constraints. First, we define two quantum circuit scheduling approaches in Sec.~\ref{sec:background}. Next, we discuss a MIP model that can schedule gates to minimize circuit execution time in Sec.~\ref{sec:methods}. Then, we analyze star graphs to determine the impact of graph structure and gate times on circuit execution time in Sec.~\ref{sec:specificgraphs}. Next, we provide simulation results in Sec.~\ref{sec:results}. Finally, in Sec.~\ref{sec:discussion}, we conclude with future research directions.

\section{Background}\label{sec:background}

The quantum circuit gate scheduling problem can be stated as follows: given a set of gates $\cG$, each with an associated execution time, and a set of precedence constraints on the gates, find an optimal schedule (circuit) that minimizes the total execution time of the circuit while respecting commutativity. 


In this work, we focus on quantum circuits with precedence constraints, where certain gates must be completed before others can begin. A prime example is the Quantum Approximate Optimization Algorithm (QAOA) and its variants \cite{farhi2014quantum, herrman2022multi, shi2022multiangle, hadfield2019quantum}, which utilize specific gate sequences with strict ordering requirements. In QAOA circuits, there are two distinct types of quantum operations: two-qubit gates (corresponding to problem edges) with execution times $\{t_{ij}^{\gamma}\}$ where $(i,j)$ represents the qubit pair acted upon, and single-qubit gates with execution times $\{t_i^{\beta}\}$ where $i$ represents the individual qubit. A critical precedence constraint in these circuits is that all two-qubit gates acting on a particular qubit must be completed before the single-qubit gate on that qubit can begin execution. Our primary objective is to minimize the total circuit execution time while respecting these precedence relationships. While we use QAOA as a motivating example, our scheduling approach applies to general circuits with arbitrary precedence constraints. Throughout this work, we will use the variables described in Tab.~\ref{tab:variables}.

\begin{table}
    \centering
    \caption{Notation}
    \begin{tabular}{|c|c|}
        \hline
        \makecell{Variable or \\ Symbol} & Description  \\
        \hline
        $t_{ij}^{\gamma}$ & \makecell{Execution time of a two-qubit gate \\acting on qubits $i$ and $j$}\\
        \hline
        $t_{i}^{\beta}$ & \makecell{Execution time of a single-qubit gate \\acting on qubit $i$}\\
        \hline
        $t_{\max}^{\beta}$ & \makecell{Maximum execution time of a \\ single-qubit gate}\\
        \hline 
        $t_{\min}^{\beta}$ & \makecell{Minimum execution time of a \\ single-qubit gate} \\
        \hline 
        $[n]$ & The set $\{0, 1, \ldots, n-1\}$ \\
        \hline
        $C(i)$ & Completion time of qubit $i$ \\
        \hline
        $\mathcal{Q}_a[i]$ & Availability time of qubit $i$ \\
        \hline
        $\pi$ & \makecell{Permutation used by greedy scheduler representing \\ the order of gates} \\
        \hline
        $\pi^{-1}(k)$ & \makecell{Pre-image of $k$ under permutation $\pi$, i.e., \\ the gate scheduled at position $k$} \\
        \hline
        $\sigma$ & \makecell{Permutation used by MIP scheduler representing \\ the order of gates} \\
        \hline
        $\sigma^{-1}(i)$ & \makecell{Pre-image of $i$ under permutation $\sigma$, i.e., \\ the position of gate $(0,i)$ in the sequence} \\
        \hline
        $\Gamma$ & Set of two-qubit gates \\
        \hline
        $\mathcal{B}$ & Set of single-qubit gates \\
        \hline
        $\cG$ & Set of all gates ($\cG = \Gamma \cup \mathcal{B}$) \\
        \hline
        $\mathcal{P}$ & Set of precedence constraints between gates \\
        \hline
    \end{tabular}
    \label{tab:variables}
\end{table}

\subsection{Representing quantum circuits as graphs}

To systematically analyze and schedule quantum circuits, we map their structure to graph-theoretic models. For circuits with two-qubit gates, we represent the circuit as a graph $G = (V, E)$ as follows:

\begin{itemize}
    \item Each vertex $v \in V$ corresponds to a single qubit in the circuit.
    \item Each edge $(i,j) \in E$ represents a two-qubit gate that acts on the qubit pair $(i,j)$.
    \item Edge weights $t_{ij}^{\gamma}$ represent the execution time of the corresponding two-qubit gate.
    \item Vertex weights $t_i^{\beta}$ represent the execution times of the single-qubit gates.
\end{itemize}

This abstraction provides a natural framework for analyzing scheduling constraints. Two gates from the gate set $\cG$ can be executed in parallel if and only if they act on distinct sets of qubits, which in the graph representation corresponds to edges with no intersecting endpoints (i.e., no shared vertices). For instance, gates represented by edges $(i,j)$ and $(k,l)$ can be executed simultaneously if and only if $\{i,j\} \cap \{k,l\} = \emptyset$.

While we focus on two-qubit gates in this work, this graph representation can be generalized to circuits with multi-qubit gates by using hypergraphs, where each hyperedge connects all qubits involved in a particular gate. 

This graph-based approach allows us to apply established graph-theoretic concepts and algorithms to the quantum circuit scheduling problem. For instance, finding sets of gates in $\cG$ that can be executed in parallel corresponds to identifying sets of non-intersecting edges in the graph, which relates to coloring problems in graph theory \cite{herrman2021lower}. This connection forms the foundation for the scheduling approaches we discuss next.

\subsection{Layered Scheduling}

The layered scheduler groups gates into ``layers," where each layer consists of gates from the gate set $\cG$ that can be executed simultaneously without resource conflicts. This approach can be conceptualized as an edge coloring problem in graph theory, where each color represents a layer and we aim to minimize the sum of maximum gate times in each color class.

In this approach, gates are typically sorted by execution time in descending order, then processed sequentially. Each gate is assigned to the earliest possible layer where it does not conflict with already assigned gates. Conflict occurs when gates share qubits, as a qubit can only be involved in one gate at a time. The execution time for each layer is determined by the longest gate in that layer, as all gates in the layer must complete before the next layer begins.

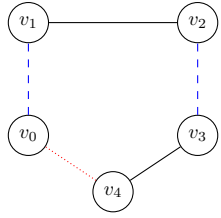
\begin{figure}
 \centering
 \begin{tikzpicture}[scale=1.5]
\begin{scope}[every node/.style={scale=.75,circle,draw}]
    \node (A) at (-.75,0) {$v_0$};
    \node (B) at (-.75,1) {$v_1$};
	\node (C) at (.75,0) {$v_3$}; 
	\node (D) at (.75,1) {$v_2$};
    \node (E) at (0,-.5) {$v_4$};

\end{scope}

\draw[blue,dashed]  (A) -- (B);
\draw   (B) -- (D);
\draw[blue,dashed]   (C) -- (D);
\draw[red, densely dotted] (E) -- (A);
\draw  (E) -- (C);

\end{tikzpicture}
\caption{A properly edge-colored cycle on five vertices, denoted $C_5$. There are three color classes: solid black, blue dashed, and red densely dotted. Note there are other minimal proper edge-colorings.}
\label{fig:maxcutproblem}
\end{figure}

To illustrate the layered scheduling approach, consider the cycle graph on five vertices, $C_5$ (see Fig.~\ref{fig:maxcutproblem}), with two-qubit gate times
\begin{align*}
&t_{01}^{\gamma} = 5, \ t_{12}^{\gamma} = 4, \ t_{23}^{\gamma} = 3, t_{34}^{\gamma} = 2, \ t_{40}^{\gamma} = 1
\end{align*}
and single-qubit gates with times $t_i^{\beta} = 1$ for all qubits $i$, where precedence constraints require that all two-qubit gates acting on qubit $i$ must complete before the single-qubit gate on qubit $i$ can begin.

The layered scheduling approach corresponds to finding a proper edge coloring of the graph. For this $C_5$ example, a valid edge coloring requires three colors (as shown in Fig.~\ref{fig:maxcutproblem}), resulting in three layers for the two-qubit gates:

\begin{enumerate}
    \item First layer: 
        \begin{itemize}
            \item Select gate $(0,1)$, $t_{01}^\gamma =5$
            \item Add gate $(2,3)$ since no conflict with $(0,1)$, $t_{23}^\gamma = 3$
            \item Layer 1 time = $\max(5, 3) = 5$
        \end{itemize}
    
    \item Second layer:
        \begin{itemize}
            \item Select gate $(1,2)$ since next longest time $t_{12}^\gamma = 4$.
            \item Add $(3,4)$ since no conflict with $(1,2)$, $t_{34}^\gamma = 2$.
            \item Layer 2 time = $\max(4, 2) = 4$
        \end{itemize}
        
    \item Third layer:
        \begin{itemize}
            \item Select gate $(4,0)$
            \item All two-qubit gates assigned to layers so layer 3 time = 1
        \end{itemize}
\end{enumerate}

After forming these layers, the single-qubit gates are scheduled in a final layer with time 1. This is because all two-qubit gates must complete before any single-qubit gate can begin, according to the precedence constraints. Note that these layers correspond to a proper coloring of the graph, as seen in Fig.~\ref{fig:maxcutproblem}.
The total execution time for the layered approach is the sum of all layer times, $t_{\text{lay}} = 5 + 4 + 1 + 1 = 11.$

In general, when precedence constraints require all two-qubit gates to complete before single-qubit gates begin, the time to execute the layered circuit is
\begin{equation*}
t_{\text{lay}} = \sum_{l \in L} \max_{g \in l} t_g
\end{equation*}
where $L$ is the set of layers.
\begin{figure*} 
    \includegraphics[width=.9\linewidth]{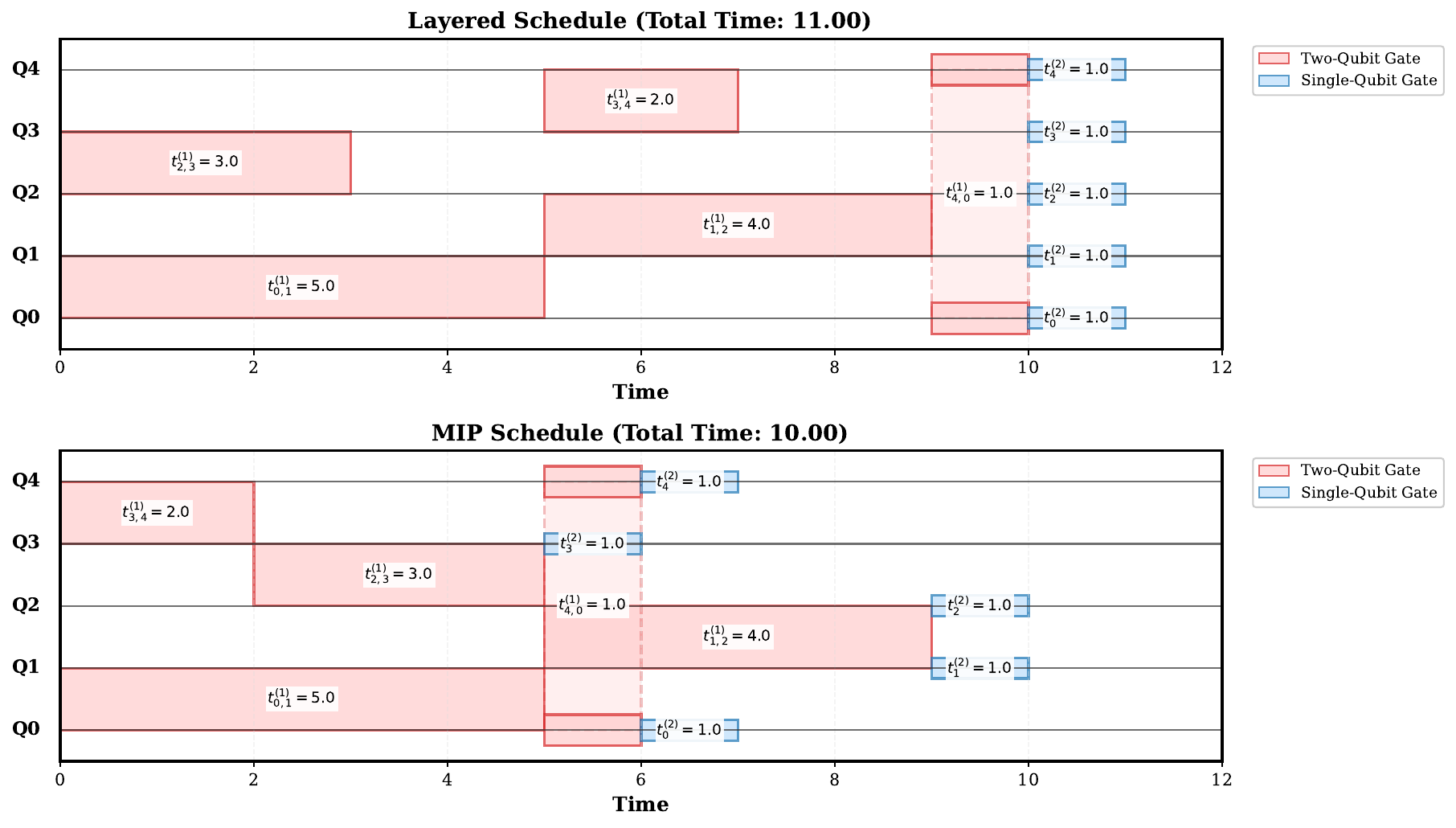}
    \caption{Comparison of scheduling approaches for the $C_5$ circuit. Top: Layered scheduling with total time 11 units. Bottom: MIP schedule with total time 10 units. The horizontal axis represents time, and each row represents a qubit. Red blocks represent two-qubit operations from the first block, and blue blocks represent single-qubit operations from the second block.}
    \label{fig:c5-schedule-comparison}
\end{figure*}

Fig.~\ref{fig:c5-schedule-comparison} shows the layered circuit schedule for the $C_5$ circuit. Each row represents a qubit, and each block represents a quantum gate. The red blocks represent two-qubit gates arranged in distinct layers, and the blue blocks represent the single-qubit gates.

The example demonstrates a key limitation of layered scheduling: even though some qubits become available earlier (qubit 3 is free after 3 time units in the first layer), the algorithm must wait until the entire layer completes (5 time units) before starting the next layer. This creates inefficiencies, particularly when gate execution times vary significantly within layers. This limitation motivates a MIP approach, which allows for more flexible scheduling.

\subsection{Greedy Scheduling}

In contrast to layered scheduling, greedy scheduling exploits the potential to start gates from the gate set $\cG$ as soon as their required qubits become available, even if other gates from the same conceptual ``layer" are still executing.
The greedy algorithm processes gates by first identifying all gates that are ready to be executed (no pending precedence constraints), then selecting the one with the earliest possible start time. This approach creates a more compact schedule by utilizing idle qubit time.

\begin{algorithm}
    \caption{Greedy Scheduling with Precedence Constraints}
    \label{algo:greedy}
    \SetAlgoLined
    \KwIn{Set of gates $\cG$ with execution times $\{t_g\}$, Set of qubits $V$, Precedence set $\mathcal{P}$}
    \KwOut{Start times $\mathcal{S}$, qubit availability $\mathcal{Q}_a$, execution time $T$}
    $\cG_{sorted} \leftarrow \text{Sort}(\cG, t_g, \text{descending})$\;
    $\mathcal{Q}_a \leftarrow \{i: 0 \mid i \in V\}$\; 
    $\mathcal{S} \leftarrow \emptyset$\;
    $\cG_{remaining} \leftarrow \cG$\;  
    
    \While{$\cG_{remaining} \neq \emptyset$}{
        $\cG_{ready} \leftarrow \{g \in \cG_{remaining} \mid \forall (g',g) \in \mathcal{P}, g' \notin \cG_{remaining}\}$\;  
        
        $\cG_{schedulable} \leftarrow \emptyset$\;
        
        \For{$g \in \cG_{ready}$}{
            $Q_g \leftarrow \text{qubits}(g)$ \; 
            $t_{start} \leftarrow \max_{q \in Q_g} \mathcal{Q}_a[q]$\;
            $\cG_{schedulable} \leftarrow \cG_{schedulable} \cup \{(g, t_{start})\}$\;
        }
        
        $(g^*, t^*) \leftarrow \text{argmin}_{(g,t) \in \cG_{schedulable}} t$\;
        $\mathcal{S}[g^*] \leftarrow t^*$\;
        
        \For{$q \in \text{qubits}(g^*)$}{
            $\mathcal{Q}_a[q] \leftarrow t^* + t_{g^*}$\;
        }
        
        \For{$g \in \cG_{ready} \setminus \{g^*\}$}{
            $Q_g \leftarrow \text{qubits}(g)$\;
            \If{$Q_g \cap \text{qubits}(g^*) = \emptyset \land \max_{q \in Q_g} \mathcal{Q}_a[q] = t^*$}{
                $\mathcal{S}[g] \leftarrow t^*$\;
                \For{$q \in Q_g$}{
                    $\mathcal{Q}_a[q] \leftarrow t^* + t_g$\;
                }
                $\cG_{remaining} \leftarrow \cG_{remaining} \setminus \{g\}$\;
            }
        }
        
        $\cG_{remaining} \leftarrow \cG_{remaining} \setminus \{g^*\}$\;
    }
    
    $T \leftarrow \max_{i \in V} \mathcal{Q}_a[i]$\;
    \Return{$(\mathcal{S}, \mathcal{Q}_a, T)$}\;
\end{algorithm}

To illustrate Algorithm \ref{algo:greedy} in action, we consider the cycle graph $C_5$ example with the same gate times as in the layered scheduling case. When executing the algorithm, the following steps occur:

\begin{enumerate}
    \item Initial setup:
        \begin{itemize}
            \item $\cG_{remaining}$ contains all two-qubit and single-qubit gates
            \item $\mathcal{Q}_a[i] = 0$ for all qubits $i$
            \item $\cG_{ready}$ initially contains only two-qubit gates, as single-qubit gates have precedence constraints
        \end{itemize}
    \item First iteration:
        \begin{itemize}
            \item Among ready gates, $(0,1)$ with time 5 and $(3,4)$ with time 2 can start at time 0
            \item Select gate $(0,1)$ and schedule it at time 0: $\mathcal{S}[(0,1)] = 0$
            \item Update availability: $\mathcal{Q}_a[0] = 5$, $\mathcal{Q}_a[1] = 5$
            \item Schedule $(3,4)$ in parallel since it doesn't conflict: $\mathcal{S}[(3,4)] = 0$
            \item Update availability: $\mathcal{Q}_a[3] = 2$, $\mathcal{Q}_a[4] = 2$
        \end{itemize}
    \item Next iteration:
        \begin{itemize}
            \item Among remaining ready gates, $(2,3)$ can start earliest at time 2 (when qubit 3 becomes available)
            \item Schedule gate $(2,3)$ at time 2: $\mathcal{S}[(2,3)] = 2$
            \item Update availability: $\mathcal{Q}_a[2] = 5$, $\mathcal{Q}_a[3] = 5$
        \end{itemize}
    \item Subsequent iterations:
        \begin{itemize}
            \item Gate $(4,0)$ with time 1: $\mathcal{S}[(4,0)] = 5$ (when both qubits are free)
            \item Update availability: $\mathcal{Q}_a[4] = 6$, $\mathcal{Q}_a[0] = 6$
            \item Gate $(1,2)$ with time 4: $\mathcal{S}[(1,2)] = 5$ (when both qubits are free)
            \item Update availability: $\mathcal{Q}_a[1] = 9$, $\mathcal{Q}_a[2] = 9$
        \end{itemize}
    \item Final iterations:
        \begin{itemize}
            \item With all two-qubit gates scheduled, single-qubit gates become ready (precedence constraints satisfied)
            \item Schedule single-qubit gates as soon as qubits become available
            \item Final qubit availability times: $\mathcal{Q}_a[0] = 7$, $\mathcal{Q}_a[1] = 10$, $\mathcal{Q}_a[2] = 10$, $\mathcal{Q}_a[3] = 6$, $\mathcal{Q}_a[4] = 7$
        \end{itemize}
\end{enumerate}

The total execution time is determined by the latest completion time among all qubits,
\begin{align*}
t_{\text{greedy}} &= \max_{q \in V} \mathcal{Q}_a[q] 
= \max\{7, 10, 10, 6, 7\} = 10
\end{align*}

This flexible approach allows for a much more efficient utilization of qubit time. For this $C_5$ example, the greedy scheduling approach results in a circuit with a shorter execution time than the layered schedule (10 units vs. 11 units), as shown in the bottom part of Fig.~\ref{fig:c5-schedule-comparison}.

\subsection{MIP Scheduling}

For the MIP scheduling approach, we use a classical optimization solver to find the optimal schedule. Our implementation uses the PuLP library in Python, which provides an interface to several commercial and open-source solvers for MIP problems. The MIP model formulation, as described in Sec.~\ref{sec:methods}, assigns start times to gates while respecting precedence constraints and resource conflicts.

Unlike the layered and greedy approaches, which use specific heuristics, the MIP solution provides a globally optimal schedule (given sufficient computational resources). The model explicitly handles the trade-offs between different scheduling decisions, taking into account the full problem structure, including gate execution times and precedence relationships.

When applied to the $C_5$ example, the MIP scheduler produces a similar schedule to the greedy approach, with a total execution time of 10 units. However, as we will show in Sec.~\ref{sec:specificgraphs}, there are cases where the MIP approach significantly outperforms both layered and greedy scheduling methods.

The total execution time for MIP scheduling is
\begin{equation*}
t_{\text{MIP}} = \min_{\mathcal{S}} \max_{g \in \Gamma \cup \mathcal{B}} \{\mathcal{S}[g] + t_g\},
\end{equation*}
where $\mathcal{S}[g]$ is the start time of gate $g$, and $t_g$ is its execution time. The minimization is over all valid schedules that respect precedence constraints and resource conflicts.

As we will show in Sec.~\ref{sec:specificgraphs}, greedy scheduling can approach or even match the performance of MIP methods for certain graph structures and gate times, while being computationally less expensive. However, for more complex circuits with varying gate times, the MIP approach can result in significantly shorter circuits than either the layered or greedy schedules.

\section{MIP Model}\label{sec:methods}

Given a set of gates $\cG$, each gate with an execution time $t_g$ for all $g \in \cG$ and a set of precedence constraints $\mathcal{P}$, we want to execute all gates in the minimum amount of time. We let $x_g$ represent the start time for gate $g$ and let $y_{g,g'}$ be a binary variable that equals 1 if gate $g$ is performed before gate $g'$, and 0 otherwise. The optimization problem is modeled as

\begin{align}
\min & \; Z \label{eq:obj}\\
\text{s.t.}\ & x_{g'} \geq x_{g} + t_{g} - My_{g',g} \quad \forall g,\ g'\ \in \cG_i,\ \forall i \in V \label{eq:c1}\\
&Z \geq x_{g} + t_{g} \quad \forall g \in \cG \label{eq:c2}\\
& y_{g,g'} + y_{g',g} = 1 \quad \forall g,\ g' \in \cG_i,\ \forall i \in V, g \neq g' \label{eq:c3}\\
& y_{g,g'} = 1 \quad \forall (g \prec g') \in \mathcal{P} \label{eq:c4}\\
& x_{g} \geq 0 \quad \forall g \in \cG \label{eq:c5}\\
& y_{g,g'} \in \{0,1\} \quad \forall g,\ g' \in \cG \label{eq:c6}
\end{align}
where $Z$ is the total execution time of the circuit, $\cG_i$ is the set of gates that operate on qubit $i$, $V$ is the set of all qubits, $\mathcal{P}$ is the set of precedence constraints, and $M$ is a sufficiently large constant.

Constraint~\eqref{eq:c1} ensures proper sequencing of gates that share qubits. This constraint implements a binary switching mechanism: when $y_{g',g} = 0$, gate $g'$ must start after gate $g$ completes, and when $y_{g',g} = 1$, the constraint becomes inactive due to the large value of $M$. This effectively creates an either-or relationship for gates that cannot run simultaneously due to shared qubits.

Constraint~\eqref{eq:c2} defines the total execution time of the circuit. It ensures that the circuit's completion time $Z$ is at least as large as the completion time of any gate. The minimization objective ensures $Z$ will equal the maximum completion time across all gates.

Constrain~\eqref{eq:c3} enforces that for any pair of gates that share a qubit, one must be executed before the other. This eliminates the possibility of simultaneous execution of gates that share resources and ensures a deterministic ordering.

Constraint~\eqref{eq:c4} enforces the precedence constraints. If gate $g$ must be performed before gate $g'$ according to the precedence constraints, this is reflected in the fixed value of $y_{g,g'}$.

Constraint~\eqref{eq:c5} ensures that all start times are non-negative, and constraint~\eqref{eq:c6} ensures that the precedence variables are binary.

This MIP model can handle quantum circuits with arbitrary gate structures and precedence constraints. For circuits with precedence requirements such as those in QAOA, where two-qubit gates must complete before single-qubit gates on the same qubit, we would include constraints $(g \prec g') \in \mathcal{P}$ for each two-qubit gate $g$ and single-qubit gate $g'$ that operate on the same qubit.

The model's flexibility allows it to find optimal schedules that minimize the total circuit execution time while respecting all constraints. Unlike the layered and greedy approaches, which rely on specific heuristics, the MIP solution guarantees optimality (given sufficient computational resources). This is particularly valuable for complex quantum circuits with varying gate times and intricate precedence relationships.

\section{Examples: Gate execution time impact on scheduling}\label{sec:specificgraphs}

We now analyze how gate execution time impacts circuit execution time for star graphs. A star graph $S_n$ has one central vertex connected to $n-1$ leaf vertices, with no edges between leaves. For quantum circuits on star graphs with two-qubit gates having times $\{t_{0j}^{\gamma}\}$ and single-qubit gates with times $\{t_i^{\beta}\}$, we can derive exact expressions for schedule times under different approaches.

Let $t_{\max}^{\gamma} = \max_{j \in [n-1]} t_{0j}^{\gamma}$, $t_{\max}^{\beta} = \max_{i \in [n]} t_i^{\beta}$, and $t_{\min}^{\beta} = \min_{i \in [n]} t_i^{\beta}$.

\subsection{Layered Scheduling}

In a star graph, all edges share the central vertex, requiring all two-qubit gates to be placed in separate layers. The total time for layered scheduling is
\begin{align*}
t_{\text{lay}} = \sum_{j=1}^{n-1} t_{0j}^{\gamma} + t_{\max}^{\beta}.
\end{align*}

This approach waits for all two-qubit gates to complete before starting any single-qubit gates, even when some qubits become available earlier.

\subsection{Greedy Scheduling}

Next, we analyze the greedy scheduling approach. Since all two-qubit gates share the center qubit (qubit 0), they must be executed sequentially. The greedy scheduler processes each gate in order, assigning the earliest possible start time based on qubit availability while respecting precedence constraints.

Initially, $\mathcal{Q}_a[i] = 0$ for all $i \in [n]$. The greedy scheduler processes two-qubit gates in some order, which we denote by a permutation $\pi$, where $\pi(j)$ indicates the position of gate $(0,j)$ in the sequence. Conversely, $\pi^{-1}(k)$ denotes the gate scheduled at position $k$.

After scheduling all two-qubit gates according to permutation $\pi$, the availability times are
\begin{align*}
\mathcal{Q}_a[0] &= \sum_{j=1}^{n-1} t_{0j}^{\gamma} \\
\mathcal{Q}_a[j] &= \sum_{k=1}^{\pi(j)} t_{0\pi^{-1}(k)}^{\gamma} \quad \text{for $j \in [n-1]$}
\end{align*}

The single-qubit gates are scheduled immediately after their precedence constraints are satisfied (all two-qubit gates acting on the same qubit have completed). The completion time $C(i)$ for each qubit $i$ is its availability time plus its single-qubit gate time,
\begin{align*}
C(0) &= \mathcal{Q}_a[0] + t_0^{\beta} = \sum_{j=1}^{n-1} t_{0j}^{\gamma} + t_0^{\beta} \\
C(j) &= \mathcal{Q}_a[j] + t_j^{\beta} = \sum_{k=1}^{\pi(j)} t_{0\pi^{-1}(k)}^{\gamma} + t_j^{\beta} \quad \text{for $j \in [n-1]$}.
\end{align*}

The total execution time for the greedy scheduler is determined by the maximum completion time across all qubits,

{\small
\begin{align*}
t_{\text{greedy}} &= \max_{i \in [n]} C(i) \\
&= \max\left\{\sum_{j=1}^{n-1} t_{0j}^{\gamma} + t_0^{\beta}, \max_{j \in [n-1]} \left(\sum_{k=1}^{\pi(j)} t_{0\pi^{-1}(k)}^{\gamma} + t_j^{\beta}\right)\right\}.
\end{align*}}

The performance of the greedy scheduler depends critically on the permutation $\pi$ it uses. Standard greedy implementations typically order gates by decreasing two-qubit gate times without considering the single-qubit gate times. This can lead to suboptimal schedules, particularly when there is significant variation in single-qubit gate times.

If a leaf qubit $i$ with a long single-qubit gate time $t_i^{\beta}$ has its two-qubit gate scheduled late in the sequence (large $\pi(i)$), it may become the critical path that determines the total execution time. Conversely, if this gate is scheduled early (small $\pi(i)$), the single-qubit gate can run in parallel with subsequent two-qubit gates that don't involve qubit $i$, potentially reducing the overall execution time.

\subsection{MIP Scheduling}

We now analyze the mixed-integer programming (MIP) approach for scheduling quantum circuits on star graphs. Using our MIP model from Section \ref{sec:methods}, we can derive precise mathematical expressions for the optimal schedule times. In practice, this MIP model is implemented using the PuLP library in Python, which interfaces with classical optimizers such as CBC, GLPK, or Gurobi to solve the optimization problem.

When we solve the MIP model for star graphs, a clear pattern emerges in the resulting schedules. For star graphs with precedence constraints requiring two-qubit gates to complete before their corresponding single-qubit gates, if all single-qubit gate times are less than the total two-qubit gate time (i.e., $t_i^{\beta} < \sum_{j=1}^{n-1} t_{0j}^{\gamma}$ for all $i \in [n-1]$), the optimal strategy orders the two-qubit gates by decreasing corresponding single-qubit gate time.

First, let $\sigma$ be the permutation that determines the order of two-qubit gates, such that gate $(0,\sigma(j))$ is the $j$-th gate to be executed. For each qubit, we calculate its availability time $\mathcal{Q}_a[i]$, which is when all precedent gates involving qubit $i$ have completed.

For the central vertex 0 which participates in all two-qubit gates, the availability time is
\begin{align*}
\mathcal{Q}_a[0] = \sum_{j=1}^{n-1} t_{0j}^{\gamma}.
\end{align*}
For a leaf qubit $k \in [n-1]$, the availability time, which depends on when its two-qubit gate completes, is
\begin{align*}
\mathcal{Q}_a[k] = \sum_{j=1}^{\sigma^{-1}(k)} t_{0\sigma(j)}^{\gamma}
\end{align*}
where $\sigma^{-1}(k)$ is the position of gate $(0,k)$ in the execution sequence.

The completion time for each qubit $i$ is its availability time plus its single-qubit gate time,
\begin{align*}
C(i) = \mathcal{Q}_a[i] + t_i^{\beta}.
\end{align*}

The total execution time is the maximum completion time across all qubits,
{\tiny
\begin{align*}
t_{\text{MIP}} = \max_{i \in [n]} C(i)  
&= \max\left\{C(0), \max_{k \in [n-1]} C(k)\right\} \\
&= \max\left\{\sum_{j=1}^{n-1} t_{0j}^{\gamma} + t_0^{\beta}, \max_{k \in [n-1]} \left\{\sum_{j=1}^{\sigma^{-1}(k)} t_{0\sigma(j)}^{\gamma} + t_k^{\beta}\right\}\right\}.
\end{align*}}

We can now manipulate the expression for a leaf qubit's completion time. Note that the sum of all two-qubit gate times can be split into two sums,
\begin{align*}
\sum_{j=1}^{n-1} t_{0j}^{\gamma} = \sum_{j=1}^{\sigma^{-1}(k)} t_{0\sigma(j)}^{\gamma} + \sum_{j=\sigma^{-1}(k)+1}^{n-1} t_{0\sigma(j)}^{\gamma}.
\end{align*}

Solving for the first term on the right side yields
\begin{align*}
\sum_{j=1}^{\sigma^{-1}(k)} t_{0\sigma(j)}^{\gamma} = \sum_{j=1}^{n-1} t_{0j}^{\gamma} - \sum_{j=\sigma^{-1}(k)+1}^{n-1} t_{0\sigma(j)}^{\gamma}, 
\end{align*}
which can be substituted into the expression for a leaf qubit's completion time, resulting in
\begin{align*}
C(k) &= \sum_{j=1}^{\sigma^{-1}(k)} t_{0\sigma(j)}^{\gamma} + t_k^{\beta} \\
&= \sum_{j=1}^{n-1} t_{0j}^{\gamma} - \sum_{j=\sigma^{-1}(k)+1}^{n-1} t_{0\sigma(j)}^{\gamma} + t_k^{\beta}.
\end{align*}

The total execution time formula can be rewritten as 
{\tiny
\begin{align*}
t_{\text{MIP}} &= \max\left\{\sum_{j=1}^{n-1} t_{0j}^{\gamma} + t_0^{\beta}, \max_{k \in [n-1]} \left\{\sum_{j=1}^{n-1} t_{0j}^{\gamma} - \sum_{j=\sigma^{-1}(k)+1}^{n-1} t_{0\sigma(j)}^{\gamma} + t_k^{\beta}\right\}\right\}
\end{align*}
}

Then, we can factor out the common term $\sum_{j=1}^{n-1} t_{0j}^{\gamma}$, resulting in
{\tiny
\begin{align*}
t_{\text{MIP}} 
&= \sum_{j=1}^{n-1} t_{0j}^{\gamma} + \max\left\{t_0^{\beta}, \max_{k \in [n-1]} \left\{t_k^{\beta} - \sum_{j=\sigma^{-1}(k)+1}^{n-1} t_{0\sigma(j)}^{\gamma}\right\}\right\}
\end{align*}}

This is our final expression for the MIP scheduling time. The formula has a clear interpretation: the term $\sum_{j=\sigma^{-1}(k)+1}^{n-1} t_{0\sigma(j)}^{\gamma}$ represents the total time of two-qubit gates that can run in parallel with qubit $k$'s single-qubit gate. This parallelism is possible because of the precedence constraints - once a qubit's two-qubit gate completes, its single-qubit gate can begin immediately, potentially overlapping with subsequent two-qubit gates that don't involve that qubit.

In the special case where all single-qubit gate times are small compared to the total two-qubit gate time, and the center qubit's single-qubit gate time isn't the minimum (i.e., $t_0^{\beta} > t_{\min}^{\beta}$), the optimal MIP scheduling time simplifies to
\begin{align*}
t_{\text{MIP}} = \sum_{j=1}^{n-1} t_{0j}^{\gamma} + t_{\min}^{\beta}.
\end{align*}
This simplification occurs because in this scenario, the leaf qubit with the minimum single-qubit gate time becomes the critical path, as its gate is scheduled last with no opportunity for parallelism with other operations.

\subsection{Comparison and Example}

Comparing the approaches, the improvement of MIP over layered scheduling in common cases is $t_{\text{lay}} - t_{\text{MIP}} = t_{\max}^{\beta} - t_{\min}^{\beta}$, which becomes significant with high variability in single-qubit gate times.

To demonstrate the advantage of MIP scheduling, consider $S_5$ with $t_i^\beta = 0.01$ for $i \in \{0,1,2,3\}$ and $t_4^\beta=1.99$, $t_{0i}^\gamma=1$ for $i \in \{1,2,3\}$ and $t_{04}^\gamma=0.01$.

With greedy scheduling ordering gates by duration, the completion time is $t_{\text{greedy}} = 5$. The MIP scheduler prioritizes the gate connected to the qubit with the longest single-qubit time, resulting in $t_{\text{MIP}} = 3.02$ - approximately 40\% faster.

This example illustrates how MIP scheduling outperforms greedy scheduling when ordering decisions significantly impact parallelism opportunities. By executing $t_{04}^{\gamma}$ first, the long single-qubit gate $t_4^{\beta} = 1.99$ can run in parallel with subsequent two-qubit gates, achieving substantial reduction in total execution time.

\section{Results}\label{sec:results}

To generalize our findings, we analyze a comprehensive dataset of connected graphs ranging from 3 to 7 vertices. We analyzed all non-isomorphic connected graphs available in the McKay graph database \cite{mckay}. Table~\ref{tab:graph-counts} summarizes the graph dataset used in our analysis.

\begin{table}[h]
\scriptsize
\centering
\begin{tabular}{|c|c|c|}
\hline
\textbf{Vertices} & \textbf{Non-isomorphic Graphs Analyzed} & \textbf{Maximum Possible Edges} \\
\hline
3 & 2  & 3 \\
4 & 6  & 6 \\
5 & 21  & 10 \\
6 & 112  & 15 \\
7 & 853  & 21 \\
\hline
\end{tabular}
\caption{Number of non-isomorphic connected graphs analyzed for each vertex count and maximum possible edges $\binom{n}{2}$.}
\label{tab:graph-counts}
\end{table}

For each graph, we assigned random gate times drawn from a uniform distribution in $(0, 2\pi]$ to simulate realistic variability in gate execution times. This angle range was chosen because QAOA and ma-QAOA angles are periodic, with period $\pi$ for $t_i^{\beta}$ (single-qubit gates) for all vertices $i$ and period $2\pi$ for $t_{ij}^{\gamma}$ (two-qubit gates) for all edges $(i,j)$. We then applied the layered, greedy, and MIP scheduling approaches to each graph instance and measured the resulting circuit execution times. The improvement percentage was calculated for each graph and aggregated by edge count, with mean and standard deviation computed for each group. 
\subsection{Layered vs. MIP}

\begin{figure*}
    \centering
    \includegraphics[width=0.9\linewidth]{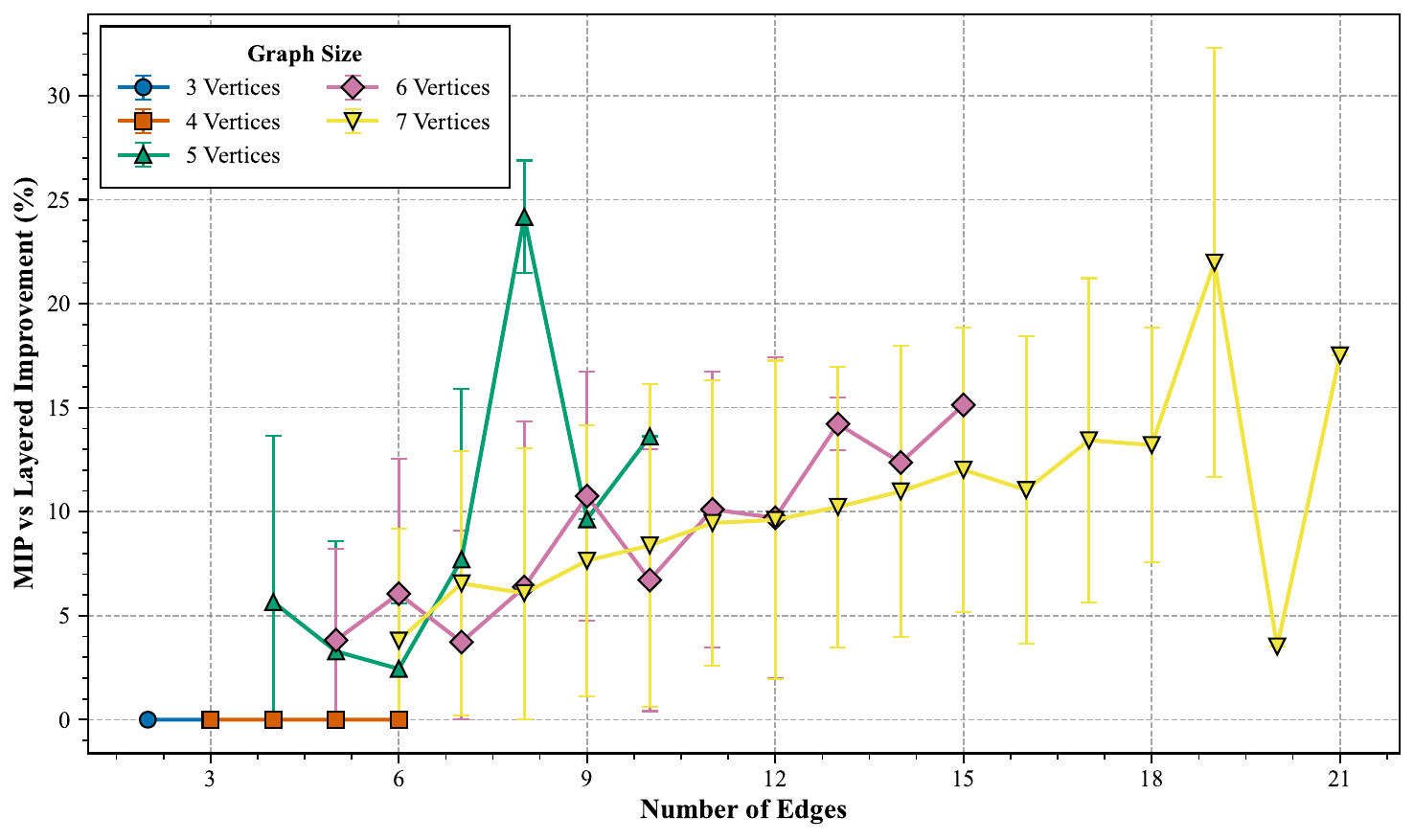}
    \caption{Improvement of MIP scheduling over layered scheduling across different graph structures with 3-7 vertices. Error bars represent standard deviation across graphs. The improvement is calculated as $\left|\frac{t_{lay} - t_{MIP}}{t_{lay}}\right| \times 100 \%$}
    \label{fig:improvement-analysis}
\end{figure*}

Our results show that graphs with 3-4 vertices produce identical schedules across all methods, with no measurable improvement. There are only two connected, non-isomorphic graph on three vertices: the path and the triangle. In both cases, the MIP scheduler produces the same schedule as the layered approach since all edges share exactly one vertex with all other edges, limiting parallelization opportunities. Similarly, the simple structures of 4-vertex graphs offer minimal scheduling flexibility.

Only graphs with 5 or more vertices show meaningful improvements. For 5-vertex graphs (maximum possible edges: 10), the peak improvement of approximately 24.2\% occurs at 8 edges (80\% of maximum edge count); for 6-vertex graphs (maximum possible edges: 15), the peak is 15.1\% at 15 edges (100\% of maximum); and for 7-vertex graphs (maximum possible edges: 21), there is a significant spike of 22\% at 19 edges (90.5\% of maximum). Fig.~\ref{fig:improvement-analysis} shows the improvement percentage of MIP scheduling over layered scheduling as a function of edge count.



\subsection{Greedy vs. MIP}

\begin{figure*}
    \centering
    \includegraphics[width=0.9\linewidth]{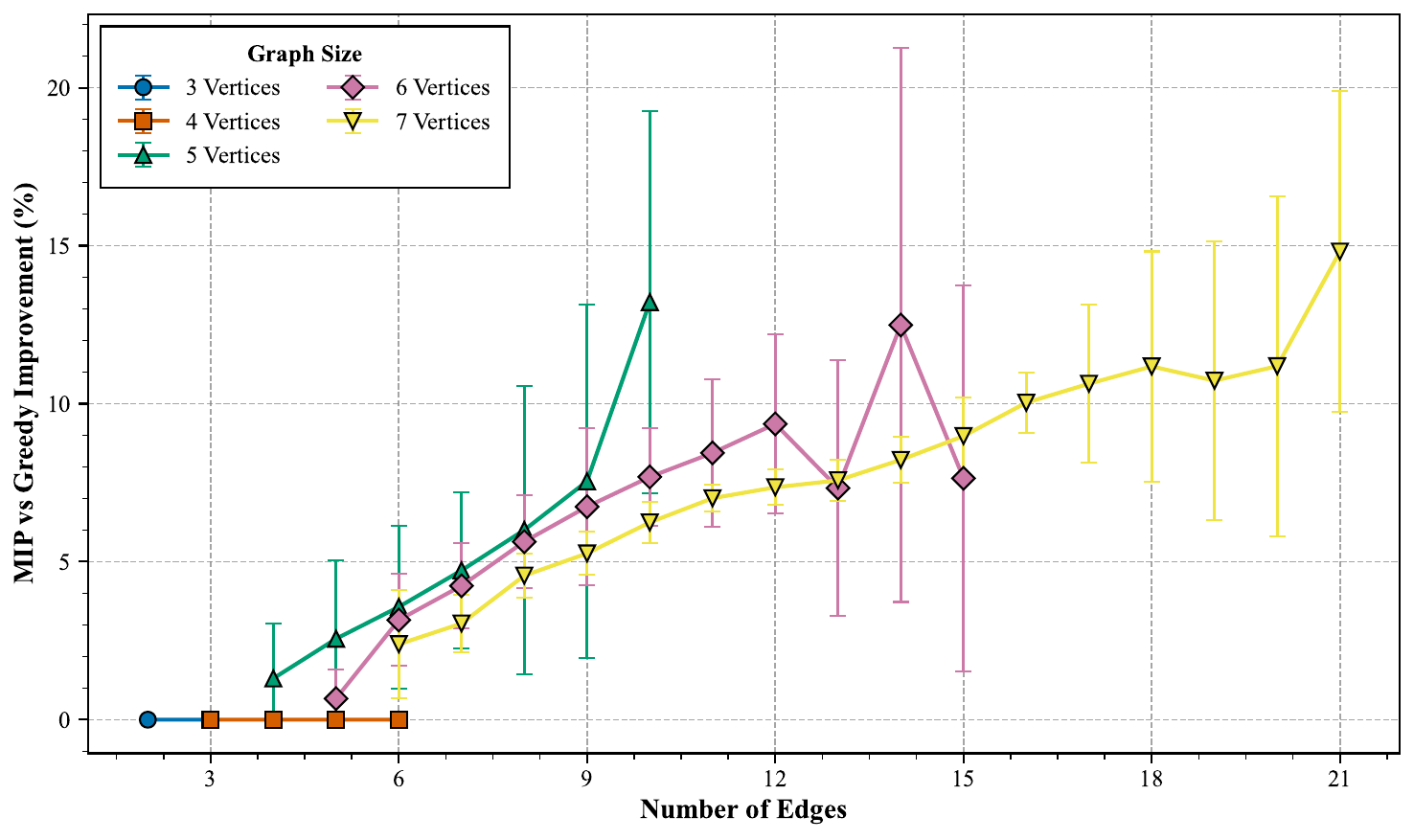}
    \caption{Improvement of MIP scheduling over greedy scheduling across different graph structures with 3-7 vertices. Error bars represent standard deviation across graphs. The improvement is calculated as $\left|\frac{t_{greedy} - t_{MIP}}{t_{greedy}}\right| \times 100 \%$}
    \label{fig:greedy-improvement-analysis}
\end{figure*}

As expected, the MIP scheduler produces an identical schedule to the greedy scheduler for 3- and 4-vertex graphs. This is due to the fact that there are few edges in these graphs, several of which share vertices, so there are few gates that can be implemented in parallel. More surprisingly, in the test graphs with random angles, the MIP schedule can show significant improvement over the greedy scheduler, in some instances resulting in approximately a 15\% improvement. 

For graphs with a fixed number of vertices between 5 and 7, the percent improvement tends to monotonically increase as the number of edges in the graphs increases, with few exceptions (6-vertex, 13 and 15 edges). These non monotonic points might be attributable to the fact that only one set of angles was generated per graph. Thus, if a particularly ``bad" set of angles was chosen, the percent improvement may be less than for other angle distributions. 

\section{Discussion}
\label{sec:discussion}
In this work, we present a new mixed-integer programming approach for scheduling quantum circuits that minimizes 
total execution time. We first compare the MIP scheduler to layered and greedy schedules for the star graph. This study exhibits that graph structure, as well as angle assignments, can greatly impact circuit execution time. In order to compare the scheduling approaches, we derived closed-form expressions for the scheduling time under each approach. Our analysis shows that the improvement of MIP over layered scheduling for star graphs is exactly $t_{\max}^{\beta} - t_{\min}^{\beta}$, which becomes significant when there is high variability in single-qubit gate times. Our theoretical analysis of star graphs provides insight into why the observed variability in improvement percentages occurs, particularly for intermediate connectivity levels. Star graphs represent a specific case that highlights how structural properties influence scheduling outcomes.

We then compare MIP schedules to layered and greedy schedules on over 950 connected, non-isomorphic graphs on up to 7 vertices. 
Figs.~\ref{fig:improvement-analysis}-\ref{fig:greedy-improvement-analysis} reveal MIP scheduling can result in significantly shorter circuit execution times compared to layered and greedy schedules, even for relatively small (5 or more vertices) problem sizes. As this study only examined one set of random angles per graph instance, future work includes running more sets of random angles on each problem instance to determine if either problem structure or angle assignment more greatly impacts the execution time of the circuit. 


While multi-angle QAOA is a motivating example and for numerical results, the MIP scheduling approach applies to any 
quantum circuit with precedence constraints, making it broadly applicable to various quantum algorithms. 
MIP scheduling exploits the fact that gates acting on disjoint qubit sets can be executed in 
parallel, allowing for significant execution time reductions compared to traditional approaches.

Future research should systematically investigate the relationship between graph theoretic properties, angle distribution, and scheduling performance to develop efficient heuristics for predicting optimal scheduling strategies. Developing predictive models that incorporate structural properties could yield powerful heuristics for scheduler selection, allowing algorithm designers to quickly estimate potential execution time improvements without solving the full MIP. 
As a motivating example, the edge chromatic number (a function of the maximum degree of a vertex in the graph) is used in layered scheduling approaches \cite{herrman2021lower}, however the relationship between graph properties and the MIP scheduler is less clear.

Additionally, extending the MIP to account for hardware-specific constraints, such as limited qubit connectivity and varying gate fidelities, could further enhance its practical impact. Integrating our scheduling approach with existing quantum compilers could provide substantial end-to-end improvements in quantum circuit execution times, potentially enabling larger and more complex algorithms to run successfully on near-term quantum hardware.

\section*{Acknowledgments}
M.A. and R.H. acknowledge DE-SC0024290. R. H. and J. O. acknowledge NSF CCF 2210063.

\section*{Competing interests}
All authors declare no financial or non-financial competing interests. 

\section*{Code and Data Availability}
\label{sec:codeAndDataAvail}
The code and data for this research can be found at \url{https://github.com/Mostafa-Atallah2020/QintProg}.

\bibliographystyle{unsrt}
\bibliography{manuscript}

\end{document}